\documentclass[preprint,showpacs,preprintnumbers,amsmath,amssymb]{revtex4}

\usepackage{graphicx}
\usepackage{dcolumn}
\usepackage{bm}
\usepackage{amssymb}

\begin{document}
\title{Comment on "Energy and information in Hodgkin-Huxley neurons"
}

\author{Hideo Hasegawa}
\altaffiliation{hideohasegawa@goo.jp;
http://sites.google.com/site/hideohasegawa/
}
\affiliation{Department of Physics, Tokyo Gakugei University,  
Koganei, Tokyo 184-8501, Japan}%

\date{\today}

\begin{abstract}
In a recent paper [A. Moujahid, A. d'Anjou, F. J. Torrealdea and F. Torrealdea, 
Phys. Rev. E {\bf 83}, 031912 (2011)], the authors have calculated the
energy consumed in firing neurons by using the Hodgkin-Huxley (HH) model. 
The energy consumption rate adopted for the HH model 
yields a {\it negative} energy consumption meaning an energy transfer
from an HH neuron to a source which is physically strange, 
although they have interpreted it as a biochemical energy cost.
I propose an alternative expression for the power consumption
which leads to a {\it positive} energy consumed in an HH neuron,
presenting some model calculations which are compared to those in their paper.

\end{abstract}

        


\maketitle
\newpage
It has been controversial how to experimentally and theoretically
evaluate the energy consumption in firing neurons.
Recently Moujahid, d'Anjou, Torrealdea and Torrealdea \cite{Moujahid11}
have proposed such a theoretical method by using the Hodgkin-Huxley (HH) model 
given by
\begin{eqnarray}
C \dot{V} &=& - I_{{\rm Na}}-I_{{\rm K}}-I_{\rm L}+I, 
\label{eq:A1} \\
\dot{m} &=& -(a_m+b_m)m+a_m, \\
\dot{h} &=& -(a_h+b_h) h+a_h,  \\
\dot{n} &=& -(a_n+b_n) n + a_n. 
\end{eqnarray}
Here $V$ denotes the membrane potential in mV, 
$I$ stands for the total current in $\mu$A/cm$^2$,
the membrane capacitance is $C=1$ $\mu$F/cm$^2$, and
$m$, $h$ and $n$ are dimensionless gating variables
of Na, K and leakage (L) channels, respectively.
Currents in respective channels are expressed by
\begin{eqnarray}
I_{{\rm Na}} &=& g_{{\rm Na}} m^3 h (V-V_{{\rm Na}}), \\
I_{\rm K} &=& g_{\rm K} n^4 (V-V_{\rm K}), \\
I_{\rm L} &=& g_{\rm L} (V-V_{\rm L}),
\label{eq:A4}
\end{eqnarray}
where reversal potentials of Na, K and L channels
are $V_{{\rm Na}}=50$ mV, $V_{\rm K}=-77$ mV and $V_{\rm L}=-54.5$ mV, respectively,
in the conventional absolute unit \cite{Hodgkin52,Note},
and the maximum values of corresponding conductances are
$g_{{\rm Na}}=120$ mS/cm$^2$, $g_{\rm K}=36$ mS/cm$^2$ and $g_{\rm L}=0.3$ mS/cm$^2$
\cite{Hodgkin52,Note}.
Coefficients of $a_m$ and $b_m$ {\it et. al.} 
are given in the absolute unit by \cite{Hodgkin52,Note}
\begin{eqnarray}
a_m &=& 0.1(V+40)/[1-e^{-(V+40)/10}], \nonumber \\
b_m &=& 4 \:e^{-(V+65)/18}, \nonumber\\
a_h &=& 0.07 \:e^{-(V+65)/20}, \nonumber\\
b_h &=& 1/[1+e^{-(V+35)/10}],\nonumber\\
a_n &=& 0.01\:(V+55)/[1-e^{-(V+55)/10}], \nonumber\\
b_n &=& 0.125 \:e^{-(V+65)/80}. \nonumber
\end{eqnarray}

In Ref.\cite{Moujahid11} the total energy $H$ is expressed by
\begin{eqnarray}
H &=& \frac{1}{2}C E^2+H_{{\rm Na}}+H_{\rm K}+H_{\rm L},
\end{eqnarray}
where the first term expresses the energy stored in the capacitor and
$H_{i}$ denotes the energy in the channel $i$ (= Na, K and L).
Based on a biochemical consideration, the authors in Ref.\cite{Moujahid11} 
have derived a derivative of $H$ with respect to time given by \cite{Moujahid11}
\begin{eqnarray}
\dot{H} &\equiv& P^{'} = CE \dot{E}+I_{{\rm Na}} E_{{\rm Na}}
+I_{\rm K} E_{\rm K}+I_{\rm L} E_{\rm L},
\label{eq:A2} 
\end{eqnarray}
where $E$ $(=V-V_{{\rm res}})$ and $E_{i}$ $(=V_{i}-V_{{\rm res}})$ 
are action potential and reversal potentials, respectively, in the reduced unit
with $V_{{\rm res}}$ $(=-65$ mV) \cite{Note}. 
The four terms in Eq. (\ref{eq:A2}) stand for energy consumption rates in
respective channels. 
By using  Eq. (\ref{eq:A2}), the authors in Ref.\cite{Moujahid11}
have studied the energy consumption rate against firing rate in an HH neuron.
Equation (\ref{eq:A2}), however, yields a negative energy consumption 
[Fig. 2(b) in Ref.\cite{Moujahid11}], for which alternative methods
are studied as will be explained in the following. 

It is well known that the HH equation given by Eq. (\ref{eq:A1}) expresses 
an electric circuit (see Fig. 1 in Ref.\cite{Hodgkin52}) 
consisting of capacitor $C$ and three Na, K and L ion channels,
which are connected in parallel. The channel $i$ (= Na, K and L)
includes a resistor $R_{i}$ and a battery $V_{i}$, through which
a current $I_{i}$ flows. The total current $I$ flows 
when these components are connected to a source battery $V$.
Applying Kirchhoff's law to the circuit, we obtain
\begin{eqnarray}
I &=& C \dot{V}+ \sum_{i} I_{i},
\label{eq:B2} \\
I_{i} R_{i} &=& V-V_{i} 
\hspace{0.5cm}\mbox{($i=$ Na, K and L)}.
\label{eq:B1}
\end{eqnarray}
Equation (\ref{eq:B2}) is nothing but Eq. (\ref{eq:A1}).
Total consumed energy rate (power) in the circuit is given by
\begin{eqnarray}
P &=& C V \dot{V} + \sum_{i} ( P_{Ji}+ P_{R i} ),
\label{eq:B4} \\
&=&  C V \dot{V} + \sum_{i} P_{i}, 
\end{eqnarray}
with
\begin{eqnarray}
P_{J i} &=& I_{i}^2 R_{i} = I_{i}(V-V_{i}),
\label{eq:C2} \\
P_{R i} &=&  I_{i} V_{i},
\label{eq:C1}
\end{eqnarray}
where $I_{i}^2 R_{i}$ and $I_{i} V_{i}$ signify contributions from Joule heat 
and reversal potential, respectively, in the channel $i$. 

Now we consider three methods A, B and C for a calculation of the power
consumed in an HH neuron, depending on which contributions are taken into account,
\begin{eqnarray}
P_A &=&  C V \dot{V}+ \sum_{i} P_{R i}
\hspace{2.5cm}\mbox{(method A)},
\label{eq:D1}\\
P_B &=&  C V \dot{V} + \sum_{i} P_{J i}
\hspace{2.5cm}\mbox{(method B)},
\label{eq:D2} \\
P_C &=&  C V \dot{V} + \sum_{i} (P_{R i}+P_{J i}) = V I
\hspace{0.5cm}\mbox{(method C)}.
\label{eq:D3}
\end{eqnarray}
Methods A, B and C take into account contributions 
from the reversal potential ($P_{Ri}$), Joule heat ($P_{Ji}$), 
and the reversal potential plus Joule heat ($P_{Ri}+P_{Ji}$),
respectively, besides that from the capacitor ($C V \dot{V}$). 
The $V I$ term in Eq. (\ref{eq:D3}) expresses a total power supplied from a source.
By a simple calculation, the total power adopted in Ref.\cite{Moujahid11} becomes 
[Eq. (\ref{eq:A2})] 
\begin{eqnarray}
P' &=& P_A - V_{{\rm res}} \:I,
\label{eq:D4}
\end{eqnarray}
which 
does not include a contribution from Joule heat.

\begin{figure}
\begin{center}
\includegraphics[keepaspectratio=true,width=100mm]{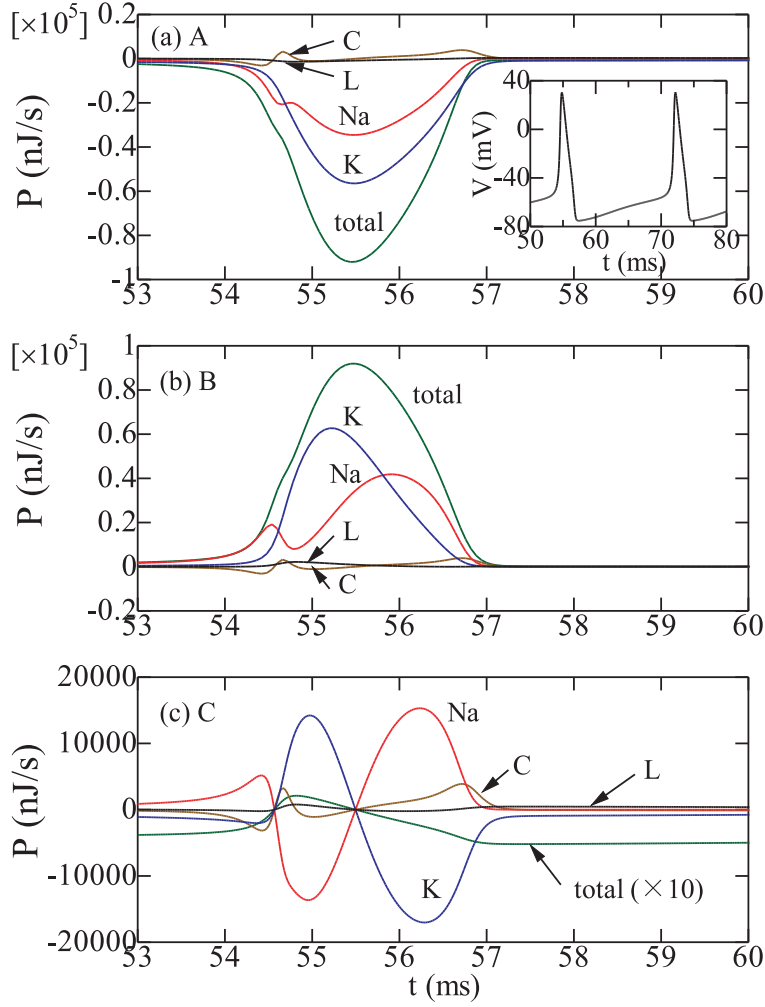}
\end{center}
\caption{
(Color online) 
Time courses of powers consumed in a capacitor (C), Na, K and L channels and
total powers
calculated in (a) the method A, (b) method B and (c) method C
with $I=6.9$ $\mu$A/cm$^2$, a total power in (c) being multiplied by a factor of ten.
The inset in (a) shows the time course of an action potential $V$.
}
\label{fig1}
\end{figure}

The inset of Fig. 1(a) shows the time course of an action potential
for an external current of $I=6.9$ $\mu$A/cm$^2$, for which
a neuron fires with a period of 17.36 ms \cite{Moujahid11}.
Properties of the consumed power depend on which method 
among the three is adopted.
Figure 1(a), 1(b) and 1(c) show time courses of power consumption in four channels
and total power for a single firing calculated in the methods A, B and C, respectively.
We note that total power $P_{A}$ in Fig. 1(a) is negative, $P_{B}$ in Fig. 1(b)
is positive, and $P_{C}$ in Fig. 1(c) is oscillating because $P_{C}=I V$.

From the total power $P$,
we may evaluate the mean total power $\bar{P}$, which is  
given by $\bar{P}_{A} < 0$, $\bar{P}_{B} > 0$ and $\bar{P}_{C} < 0$
for methods A, B and C, respectively,
a bar denoting an average over a period.
Absolute mean powers $\vert \bar{P}_{\kappa} \vert$ ($\kappa$ = A, B and C)
calculated in the methods A, B and C are
plotted by solid, dashed and chain curves, respectively, 
as a function of external current $I$ in Fig. 2, whose inset shows
the $I$ dependence of firing frequency $f$ of an HH neuron.
The mean power of firing state in the method A is about 10000 $\sim$ 15000 nJ/s 
for $I=$ 7 $\sim$ 30 $\mu$A/cm$^2$ while
that of the quiescent state without firings is much smaller (300 $\sim$ 900 nJ/s).
$\vert \bar{P}_A \vert$ shown by the solid curve is similar to the result presented 
in Fig. 3(b) of Ref.\cite{Moujahid11}.
The mean power in the method B is almost the same as that in the method A, 
\begin{eqnarray}
\bar{P}_{B} 
&\simeq& \vert \bar{P}_{A} \vert, 
\label{eq:D5}
\end{eqnarray}
which arises from the relation,
\begin{eqnarray}
\sum_i P_{Ji} &=& - \sum_i P_{R i}+ V I \simeq - \sum_i P_{R i},
\label{eq:D6}
\end{eqnarray}
because $V I \ll \vert \sum_i P_{R i} \vert$ (Fig. 1).
The mean power in the method C is smaller than those in the methods A and B,
and it is almost linearly increased with increasing $I$.
The positive mean power implies that energy is supplied from an external source to a neuron 
while the negative power means the opposite in which
a neuron plays a role of a power generator. 
The authors in Ref. \cite{Moujahid11} have interpreted an obtained negative energy 
consumption rate ($\overline{P'} < 0$) as a loss in electrochemical energy which 
is expected to correspond to consumed power because it has to be recharged 
by adenosine triphosphatase (ATPase) activity.
Equation (\ref{eq:D6}) shows that the energy consumption assumed in the method A 
(also in Ref.\cite{Moujahid11}) may be related to Joule heat
consumed in conductors of ion channels in the model B.

\begin{figure}
\begin{center}
\includegraphics[keepaspectratio=true,width=100mm]{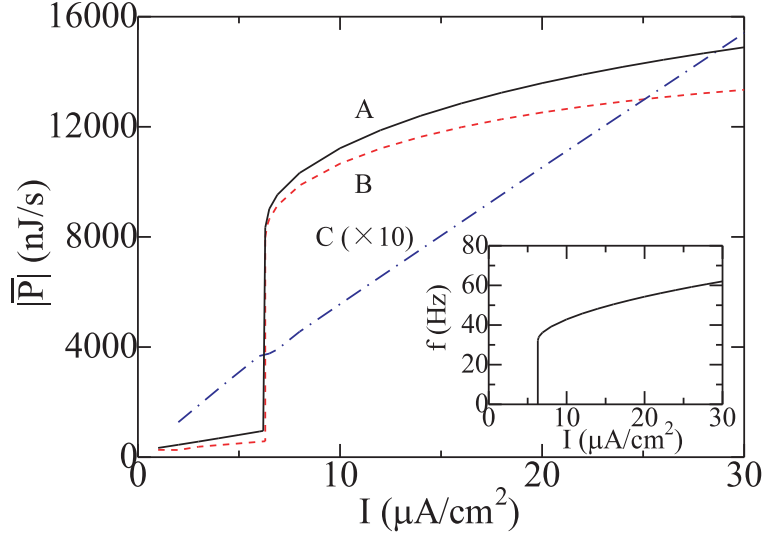}
\end{center}
\caption{
(Color online) 
Absolute mean total power $\vert \bar{P} \vert$ as a function of input current $I$
calculated in the method A (solid curve), method B (dashed curve) and method C (chain curve),
result of the method C being multiplied by a factor of ten.
The $I$ dependence of firing rate $f$ is shown in the inset where
a firing occurs at $I > 6.2$ $\mu$A/cm$^2$.
}
\label{fig2}
\end{figure}

The authors in Ref.\cite{Moujahid11} have extended their analysis
to two-electrically coupled
HH neurons and also to a collection of uncoupled HH neurons,
for which expressions for energy consumption rates similar to Eq. (\ref{eq:A2}) 
have been derived ({\it e.g.} Eq. (8) in Ref. \cite{Moujahid11} for two-HH neurons).
They have the same deficit as in the case of a single HH neuron
having been mentioned above. 

To summarize, the calculation in Ref.\cite{Moujahid11}
based on Eq. (\ref{eq:A2}) yields a negative mean power consumption in an HH neuron, 
which is physically inappropriate, although its biochemical interpretation 
is reasonable.
We have proposed an alternative method, considering three methods A, B and C 
given by Eqs. (\ref{eq:D1})-(\ref{eq:D3}).
The method B which takes into account a contribution from Joule heat consumed 
in conductors, leads to a positive mean power in the HH model.
Energies calculated in the reduced unit adopted in 
Ref. \cite{Moujahid11} are different from those in the absolute unit \cite{Note}. 
For example, the mean value of $\overline{E I}$
in the reduced unit is positive \cite{Moujahid11} while its counterpart
$\overline{V I}$ in the absolute unit is negative. 
By using the method B,
we may study nonequilibrium properties of an HH neuron
from a viewpoint adopted in Ref.\cite{Zon04} where an analogy 
between an electric RC circuit and a damped Brownian particle is employed.



\begin{thebibliography}{99}
\bibitem{Note}
Ref. \cite{Moujahid11} has adopted the reduced unit
in which the membrane potential $E$ and reversal potentials 
$E_{i}$ ($i=$ Na, K and L) are given by
$E=V-V_{{\rm res}}$ and $E_{i}=V_{i}-V_{{\rm res}}$ ($i=$ Na, K and L)
where $V_{{\rm res}}$ $(=-65$ mV) denotes the resting action potential
in the absolute unit.
In the reduced unit, the resting action potential vanishes and reversal potentials
are $E_{{\rm Na}}=115$ mV, $E_{\rm K}=-12$ mV and $E_{\rm L}=106$ mV \cite{Moujahid11}.
\end{thebibliography}
\end{document}